\pdfoutput=1
\documentclass[iop]{emulateapj}
\usepackage{hyperref}

\usepackage[usenames,dvipsnames]{color}
\usepackage{rotating}
\usepackage{epstopdf}
\usepackage{color}
\usepackage{multirow}
\usepackage[para,online,flushleft]{threeparttable}

\newcommand{\lsim}{\raise0.3ex\hbox{$<$}\kern-0.75em{\lower0.65ex\hbox{$\sim$}}}

\newcommand{\msun}{M$_{\odot}$}
\newcommand{\pom}{\,$\pm$\,}

\newcommand{\kms}{km\ s$^{-1}$}

\newcommand{\HI}{\mbox{\normalsize H\thinspace\footnotesize I}}

\usepackage{fancyheadings}
\usepackage{amsmath}
\usepackage{amssymb}
\usepackage{subfigure}
\usepackage{graphics}
\usepackage{comment}
\usepackage{natbib}
\bibpunct{(}{)}{;}{a}{}{,}
\usepackage{appendix}

\begin{document}

\title{Tidally Induced Offset Disks in Magellanic Spiral Galaxies}

\author{Stephen A. Pardy} 
\affil{Department of Astronomy, University of Wisconsin, 475 North Charter
  Street, Madison, WI 53706, USA} 
\email{spardy@astro.wisc.edu}

\author{Elena D'Onghia\altaffilmark{1}}
\affil{Department of Astronomy, University of Wisconsin, 475 North Charter
  Street, Madison, WI 53706, USA} 
\author{E. Athanassoula} 
\affil{Aix Marseille Universit\'e, CNRS, LAM (Laboratoire d'Astrophysique de Marseille) UMR 7326, 13388,
Marseille, France} 
\author{Eric M. Wilcots}
\affil{Department of Astronomy, University of Wisconsin, 475 North Charter
  Street, Madison, WI 53706, USA} 
  \author{Kartik Sheth}
  \affil{National Aeronautics and Space Administration Headquarters, 300 E. Street SW, Washington, DC 20546}

\altaffiltext{1}{Alfred P. Sloan Fellow}

\begin{abstract}

Magellanic spiral galaxies are a class of one-armed systems that often exhibit  
an offset stellar bar, and are rarely found around massive spiral galaxies. 
Using a set of N-body and hydrodynamic simulations 
we consider a dwarf-dwarf galaxy interaction as the driving mechanism for the formation of this 
peculiar class of systems. We investigate here the relation between the dynamical, stellar and 
gaseous disk center and the bar. In all our simulations the bar center always coincides with the 
dynamical center, while the stellar disk becomes highly asymmetric during the encounter causing the photometric center of the Magellanic galaxy disk 
to become mismatched with both the bar and the 
dynamical center. The disk asymmetries persist for
almost 2 Gyr, the time that it takes for the disk to be
re-centered with the bar, and well after the companion has passed. 
This explains the nature of the offset bar found in many Magellanic-type galaxies, 
including the Large Magellanic Cloud (LMC) and NGC 3906. In particular, these results, once applied to the LMC, 
suggest that the dynamical center should reside in the bar center instead of
the \HI\ center as previously assumed, pointing to a variation in the current
estimate of the north component of the LMC proper-motion.

\end{abstract}

%--------------------------
%INTRODUCTION
%--------------------------
\section{Introduction}
\label{sec:intro}

Bars are common features in today's Universe. Over two-thirds of all nearby disk galaxies are
barred (Sa-Sd) galaxies, including the Milky Way (e.g. \citealt{deVaucouleurs:1963kx, Eskridge:2000cz, 
MenendezDelmestre:2007fl, Marinova:2007dp, Sheth:2008bd}). 
The Spitzer Survey of Stellar Structure in Nearby Galaxies, S$^4$G, a well-suited survey to identify bars, confirmed 
these results for the nearby spiral galaxies  \citep{Sheth:2010jh, MunozMateos:2013jm, Buta:2015hm}. 

However, there is a class of low-mass stellar galaxies, named
Magellanic spirals, that show evidence of relatively rare
features. In particular, these galaxies are characterized by a bar whose center is displaced from that of the disk, one-armed spirals and an otherwise normal and gas-rich disk \citep{deVaucouleurs:1972ef}. The Large Magellanic Cloud (LMC) is considered the prototype of this class
of objects. However, despite a wealth of data, there is still a good
deal of uncertainty concerning the nature of the LMC's
bar (see \citet{DOnghia:2015tc} for a review on the subject). 
Work by \citet{vanderMarel:2001cr} found that the bar is offset from the
dynamical center of the LMC and resides within a large stellar disk.
Earlier work by \citet{Zhao:2000cx} described the bar as being an
unvirialized structure that is offset from the rest of the disk as a
result of the LMC's interaction with the Small Magellanic Cloud
(SMC). \citet{Subramaniam:2009vg} used the OGLE III survey
\citep{Udalski:2008um} and contend that the LMC's bar resides in the
plane of the disk. Lastly, there is no evidence of a bar in the
\HI\ maps of the LMC presented in \citet{StaveleySmith:2003bl}.

The dynamics, structure, and star formation history of the LMC have long been interpreted in the context of its
proximity to both the Milky Way and the SMC. The majority of the
observed Magellanic spirals in the nearby Universe share the LMC's
structure, in particular the evidence of an offset bar and a one-armed
spiral structure, but are rarely found
around massive spirals \citep{Wilcots:2004gv}. A good example of these systems is 
offered by NGC 3906 that shows evidence of the bar
offset from the photometric center of the galaxy by 1.2 kpc (see, e.g., Figure 1 of \citealt{deSwardt:2015fk}), or NGC 4618
\citep{1991AJ....101..829O}.  As more galaxies have been examined from
the S$^4$G survey \citep{Sheth:2010jh}, it has been found that these offset bars, though
rare, represent as many as 5\% of all barred spirals
\citep[][Sheth, K. et al. 2016, in preparation]{2012AAS...21941706R}.

The dynamics of Magellanic barred spirals is much more complex
than that of standard galaxies with centered bars, as could be expected since the
different components, disk and bar, have different centers. \citet{deVaucouleurs:1972ef} studied the orbital
structure of such galaxies using an axisymmetric and a bar potential whose centers do
not coincide. \citet{Colin:1989tc} calculated the corresponding shifts of
the Lagrangian points, as well as the response density and velocity
fields in such potentials. They found that the center of the velocity
field does not necessarily coincide either with the center of the bar or with that of the 
axisymmetric component, but may be at an intermediate position, its
specific location depending on the geometry and size of the
offset. 

N-body simulations allowed for self-consistent modeling and
the introduction of a perturbing companion. 
\citet{Athanassoula:1996} reproduced the one-armed morphology and noted
the importance of the impact position with respect to the bar, while
\citet{Bekki:2009ij} discussed whether the offset observed in the LMC
could be due to a `dark' companion. \citet{Athanassoula:1997tx} found
that the displacements of the centers are accompanied by changes of
the bar pattern speed and size. This work was extended by
\citet{Berentzen:2003dw} who introduced gas in the simulation and found
that it is possible to destroy the bar while keeping the disc structure.
More recently, numerical experiments
suggest a dwarf-dwarf galaxy interaction origin for the offset bar and one-arm
structure \citep{Besla:2012jc, Yozin:2014wc}. 
The tidal interactions between dwarf galaxies with different masses play a key role in
the stellar and gas stripping in low-mass systems as a consequence of resonant interactions between spinning discs
\citep{DOnghia:2009kf, DOnghia:2010fk, Lokas:2015wu, Gajda:2015ti} and explain the morphology 
and origin of the Magellanic Stream \citep{Besla:2010kg, Besla:2012jc}. 

In this study we aim to understand the structure and dynamics
of Magellanic spirals with similar characteristics to the LMC.  
In particular, we explore whether direct dwarf-dwarf collisions with a mass ratio of 1:10 
produce the asymmetric structures characteristic of Magellanic-type
galaxies and can account for the internal morphology and kinematics of this general class
of galaxies. While previous works showed that these systems can have
offset bars originated by encounters with a less massive companion, here 
we show that the bar is never off-centered. Instead, the stellar disk is 
shifted from the dynamical center as a consequence of the impact with a
companion galaxy.  

The paper is organized as follows. 
In \autoref{sec:methods}, we outline our methodology and describe
numerical models involving the direct collision between a Magellanic galaxy and a companion.
We then analyze the outcome of the numerical experiments in \autoref{sec:results}. A comparison of the 
models and the observations is also presented in \autoref{sec:discussion}. We then
discuss the implications to the estimate of the LMC proper-motion and conclude with a summary.

%--------------------------
%METHODS
%--------------------------
\section{Numerical Methods}
\label{sec:methods}
\begin{figure}[htbp] %  figure placement: here, top, bottom, or page
   \centering
   \includegraphics[width=3in]{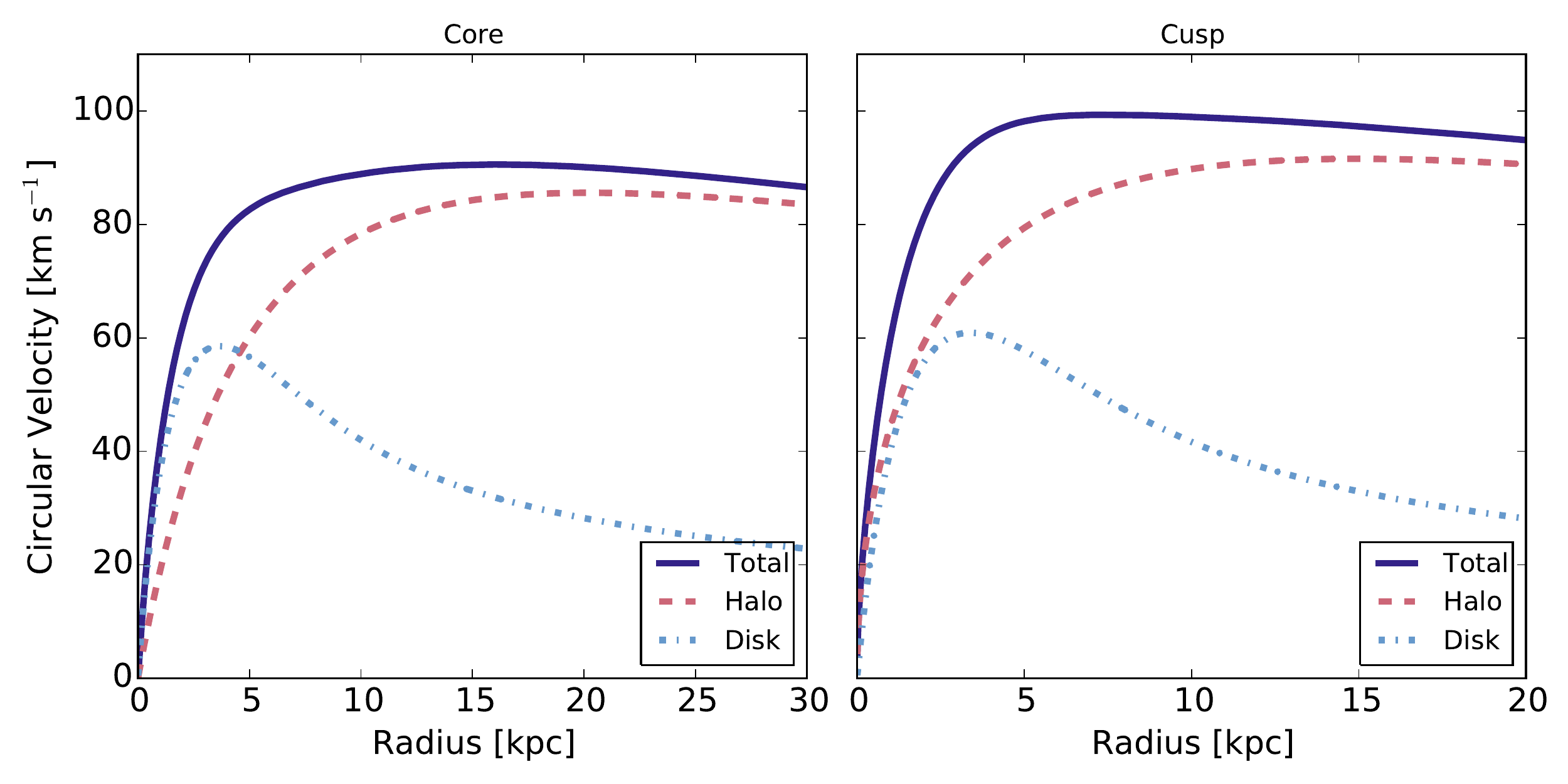} 
   \caption{Rotational curves for the primary galaxy when a cored dark profile
     (left panel) or Hernquist (right panel) is adopted.
     Both galaxies have the same total mass. A thin exponential
     disk for both the stars and gas is assumed, with the combined profile shown (light-blue dot-dash line).}
   \label{fig:rot_curves}
\end{figure}

We carried out a set of simulations with GADGET3, a parallel
TreePM Smoothed particle hydrodynamics (SPH) code developed to compute
the evolution of stars and dark matter, which are treated as
collisionless fluids and gas. A detailed description of the code is available
in the literature \citep{Springel:2005cz}. GADGET3 computes
the short-range forces using a tree-based hierarchical
multipole expansion. Pairwise particle interactions are softened with
a spline kernel of scale-length $h_s$, so that they are strictly
Newtonian for particles separated by more than $h_s$. The resulting
force is roughly equivalent to traditional Plummer softening with
scale length $h \approx h_s/2.8$.  For our applications the
gravitational softening length is fixed to $h_s=120$ pc throughout the
evolution of the galaxy encounters.

The GADGET3 code incorporates a subresolution multiphase model of the interstellar medium (ISM) including radiative cooling
\citep{Springel:2003eg} and a fully conservative
approach to integrating the equations of motion \citep{Springel:2002ef}. 
The simulations presented in this work include gas physics but are not aimed at describing properties of star formation or feedback.

\subsection{Initial Conditions}
\label{subsec:ICs}

The setup of each galaxy in our study consists of a dark matter halo and a rotationally supported
disk of stars and gas. The parameters describing each component are independent, and models are
chosen with orbital parameters as described below.

Each disk galaxy was generated using the GalIC code
described in \citet{Yurin:2014db}, which uses an iterative approach for the
realization of equilibrium N-body systems for given density distributions \citep{Rodionov:2009bo}.
 This is achieved for arbitrary axisymmetric density structure
and by taking the mutual influence of different mass components fully into account.

The initial conditions for the setup of  a Magellanic-type galaxy  and a companion
galaxy used for all models are summarized in Table 1. The total mass for the
Magellanic spiral is assumed to be an LMC analog, with parameters with values as
in \citet{Besla:2012jc}. The companion galaxy is then chosen to
be 10 times less massive than the primary galaxy. 

The number of particles of each component (gas, stars, dark matter) is chosen such that the mass
resolution per particle of a given type is roughly the same in both galaxies. 
The primary Magellanic-type galaxy  is modeled with gas and stellar disks with the same
scale length, with the values indicated in \autoref{tab:ic_params}.

% Requires the booktabs if the memoir class is not being used
\begin{table}[htbp]
   \centering
\caption{Structural Parameters of the Primary and Companion Galaxy} % requires the topcapt package
      \begin{threeparttable}
   \begin{tabular}{@{} lccr @{}} % Column formatting, @{} suppresses leading/trailing space
      \hline
      & \multicolumn{2}{c}{Primary Galaxy} &  Companion \\
      %\cmidrule(r){2-4} % Partial rule. (r) trims the line a little bit on the right; (l) & (lr) also possible
      Parameters    & Cored Halo &  Hernquist Halo &   \\
      \hline
                 
      a (kpc)\tnote{a}    & 11.3 & 14.9  & 5.8 \\
      r$_d$ (kpc) \tnote{b}     & 1.6  & 1.6  & 0.6 \\
%      C       &   15 & 15 & 15 \\
      M$_{DM} (\times 10^{10}$ M$_{\odot}$)\tnote{c} & 12   & 12  & 1.1025 \\
      M$_{disk} (\times 10^{10}$ M$_{\odot}$)\tnote{d} & 2.52 & 2.52 & 0.055 \\
      M$_{gas} (\times 10^{10}$ M$_{\odot}$)\tnote{e} & 1.08 & 1.08 & ... \\
      f$_{gas}$\tnote{f} & 0.3 & 0.3 & ...\\
      N$_{halo}$\tnote{g} & 2$\times10^6$ &2$\times10^6$ & 3.61$\times10^5$ \\
       N$_{disk}$\tnote{h} & 7$\times10^5$ & 7$\times10^5$ & 3$\times10^5$\\
       N$_{gas}$\tnote{i} & 3$\times10^5$ & 3$\times10^5$ & ...\\
      \hline
   \end{tabular}
          \begin{tablenotes}
        	   \item[a] Dark matter halo scale length.
	   \item[b] Stellar and gaseous disk scale length.
	   \item[c] Mass in dark matter.
	   \item[d] Mass in stars.
	   \item[e] Mass in gas.
	   \item[f] Baryonic gas fraction.
	   \item[g] Number of dark matter particles.
	   \item[h] Number of stellar particles.
	   \item[i] Number of gas particles.
        \end{tablenotes}
     \end{threeparttable}

%\caption{Structural parameters of the primary galaxy and the companion.}
   \label{tab:ic_params}
\end{table}

The disk component in each interacting galaxy has a thin exponential
surface density profile of scale length $r_{d}$:

\begin{equation}
\Sigma_{disk}=\frac{M_{disk}}{2\pi r_{d}^2}\exp(-r/r_{d}),
\end{equation}

\noindent
so that the disk mass is M$_{disk} = m_{d}$M$_{\text{DM}}$, where $m_{d}$ is dimensionless and M$_{\text{DM}}$ is the total
halo mass.  
The vertical mass distribution of the stars in the disk is specified by giving it the profile
of an isothermal sheet with a radially constant scale height $z_{0}$. The 3D stellar density in
the of stars and gas disk is hence given by:

\begin{equation}
\rho_*(r,z)= \frac{M_{disk}}{4\pi z_{0}r_{d}^2} \rm{sech}^2 \Big(\frac{z}{z_0}\Big) exp\Big(-\frac{r}{r_{d}}\Big)
\end{equation}
The scale height of the stellar disk is adopted as 0.2 of the disk
scale length.  The gaseous disk height is initially set equal to the stellar disk height. The energy and pressure
of the ISM are prescribed by the chosen
effective equation of state \citep{Springel:2005co}.

The models for the primary Magellanic-type galaxy assume a dark halo
with the following general form \citep{Dehnen:1993ts}:

\begin{equation}
\rho(r) = \frac{(3-\gamma)M_{DM}}{4\pi}\frac{a}{r^{\gamma}(r+a)^{4-\gamma}},
\end{equation}  
where $M_{\text{DM}}$ is the galaxy halo mass, $a$ is the scale length of the halo, and $\gamma$
is a parameter that determines the shape of the profile. For $\gamma = 1$ the
halo has a central cusp and follows a Hernquist model \citep{Hernquist:1990hf}, while for $\gamma = 0$ 
the density profile belongs to the same family of density profiles but with a
constant-density core. The isotropic distribution function for
the energy is given in this case by \citep{Dehnen:1993ts}:

\begin{equation}
f(\epsilon) = \frac{(3-\gamma)M}{2(2\pi^2GMa)^{3/2}}
\int_0^{\epsilon}\frac{(1-y)^2[\gamma +2y+(4-\gamma)y^2]}{y^{4-\gamma}\sqrt{\epsilon-\Psi}}d\Psi 
\end{equation}
where $\epsilon = -E(GM/a)^{-1}$ is the dimensionless binding energy. 

The total galaxy mass and disk mass fraction for the primary galaxy
are fixed to be the same in both models. Additionally, the
scale lengths for the cored and Hernquist halo models are set such that
both galaxies have similar rotation velocity $V_{\text{tot}}$ at the radius that enclosed
the total mass of the galaxy.

\autoref{fig:rot_curves} displays the resulting rotation curves for
the primary Magellanic-type galaxy when a cored density profile (left
panel) or a Hernquist profile (right panel) is adopted. The initial
total rotation curve of the primary galaxy peaks at $V_{\text{rot}}$ = 90
\kms\ at two halo scale lengths from the center in the cored profile,
whereas the peak is 100 \kms\ for the cusped Hernquist model at one
halo scale length. This gives a total mass within 8 kpc for the cored
(cusped) galaxy of 1.35 (1.77) $\times$10$^{10}$M$_{\odot}$. Note that the disk
fraction within 2$r_d$ is higher in the cored galaxy with respect to the
Hernquist galaxy.

\begin{figure*}[htbp] %  figure placement: here, top, bottom, or page
   \centering
   \includegraphics[width=7in]{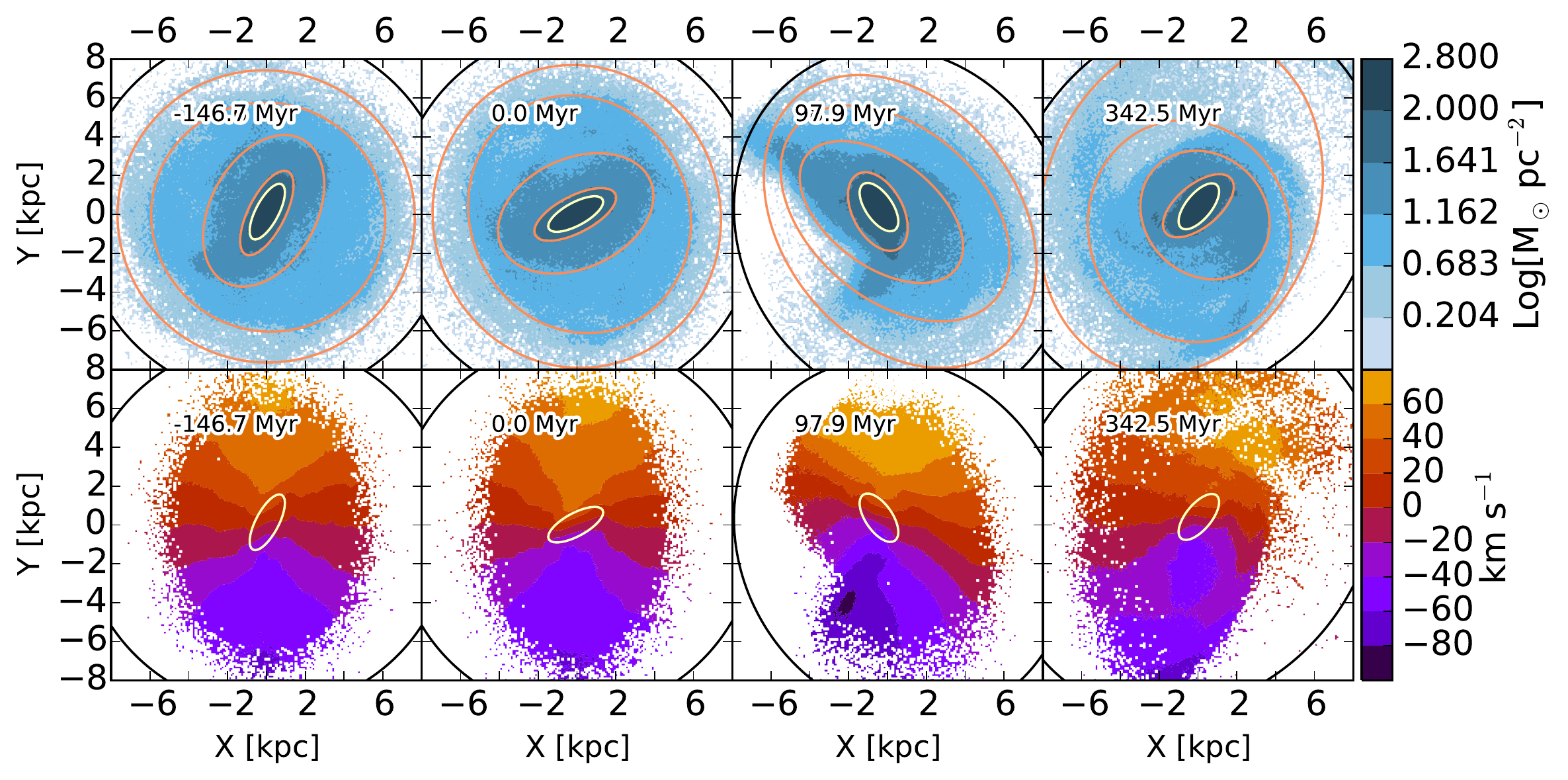} 
   \caption{Time sequence of the stellar and gaseous disks of the primary galaxy. The leftmost panel corresponds to the time immediately before the companion is introduced. Time 0 ($t$ = 0) is set at the time 
of the closest approach with the perturber.  The primary galaxy is modeled here with a cored halo density model. 
Note that the stellar and gas distributions begin orderly but become disturbed during the encounter with the companion. The top row shows 
   the face-on stellar disk surface density
     map. Isodensity ellipses (orange
     lines) are superimposed. The inner ellipse is used to identify the bar (yellow),
     while the outer ellipse outlines the disk (black). The
     encounter in this model is prograde with an inclination of 45$^{\circ}$. The companion is not
     displayed in the image sequence. The bottom row displays the time evolution of the gas velocity fields of the same galaxy. The effect of the bar is apparent in the first panel as an asymmetry in the inner contours. }
   \label{fig:combinedMaps}
\end{figure*}

This study focuses on the response of the disk of a
Magellanic-type galaxy when perturbed by a close encounter with a
lower-mass companion. The initial stellar disks of the Magellanic-type
galaxies are chosen to be sufficiently bar unstable for the bar to grow fast in isolation \citep{DOnghia:2015bi}; hence, the bar feature in all models has not been induced by external tidal perturbations 
induced by the companion \citep{Lokas:2014ji}, but grew from the beginning of the simulation.

The perturber is modeled assuming an exponential stellar disk and a Hernquist dark matter density profile, with the choice of parameters listed in \autoref{tab:ic_params}.  

\subsection{Orbital configurations}
\label{subsec:orbits}
Our set of simulations presents four orbital configurations for
each halo model.  In each set of simulations we let the primary galaxy
evolve over a period of 4 Gyr in order to grow a stellar bar at the
center and to settle to equilibrium before we introduced a
companion. At the start of the simulation, the companion was placed at
$\sim$ 50 kpc on a fast single passage orbit. Visual inspection of the primary disk 
after the companion has been added shows no perturbations until the impact occurs.
The initial position and velocities of the companion galaxy in the set of
simulations are presented in \autoref{tab:orbits}. The values are assumed in
the coordinate system at the center of mass of the primary galaxy. 
The initial orbital position and velocity of the interacting galaxies  are set so that all encounters had an impact
parameter of 4 \pom\ 0.06 kpc and a relative velocity of 337 \pom\ 4
\kms. This value is motivated by simulations of the LMC in which an offset bar is created by a direct impact by the SMC \citep{Besla:2012jc}.

% Requires the booktabs if the memoir class is not being used
\begin{table}[htbp]
   \centering 
   \caption{ORBITAL PARAMETERS OF NUMERICAL MODELS.}
      \begin{threeparttable}
   \begin{tabular}{@{} lcccccr @{}} % Column formatting, @{} suppresses leading/trailing space
      \hline
      Model & x & y & z & Vx & Vy & Vz \\
                 & \multicolumn{3}{c}{(kpc)} & \multicolumn{3}{c}{(\kms)} \\
      \hline
      Cored, $\theta=45^{\circ}$ & 29 & 36 & 29 & -147 & -147 & -147 \\
      Cored, $\theta=90^{\circ}$  & 0 & 6 & 50 & 0 & 0 & -255 \\
      Cored, $\theta=0^{\circ}$ & 50 & 7 & 0 & -260 & 0 & 0 \\
      Cored, $\theta=0^{\circ}$ Retrograde & 50 & 7 & 0 & -265 & 0 & 0 \\
      Hernquist, $\theta=45^{\circ}$ & 29 & 35 & 29 & -147 & -147 & -147 \\
      Hernquist, $\theta=90^{\circ}$  & 0 & 3.75 & 50 & 0 & 0 & -259 \\
      Hernquist, $\theta=0^{\circ}$ & 50 & 3.5 & 0 & -255 & 0 & 0 \\
      Hernquist, $\theta=0^{\circ}$ Retrograde & 50 & 3.5 & 0 & -255 & 0 & 0 \\
      \hline
   \end{tabular}
           \begin{tablenotes}
        	  Note: The origin of the coordinate system is at the center-of-mass of the primary galaxy.
        \end{tablenotes}
     \end{threeparttable}
   \label{tab:orbits}
\end{table}

Two out of four orbital configurations are coplanar 
($\theta$ = 0). One is a prograde, and the other is retrograde. The other two orbits are
both prograde and inclined with respect to the plane of the
primary galaxy by 45$^{\circ}$ and 90$^{\circ}$, respectively. The encounter occurred
at $\sim$ 0.15 Gyr and the primary galaxy was followed for 4 Gyr
after the simulation started.

\subsection{Quantifying disk off-centers and lopsidedness}
\label{subsec:centers}

In order to quantify the bar displacement in the Magellanic-type galaxy in tidal interaction with a companion galaxy, we first need to identify the centers of the galaxy components: stellar and gas disks and the stellar bar.  
An observationally motivated approach has been used to define disk and bar
centers as in the study of the Magellanic spiral NGC 3906 by \citet{deSwardt:2015fk}.

The stellar or gaseous disk of the primary galaxy is projected to a face-on density
map with 256 square bins in each dimension. Each bin has a size of $\sim$78
pc, corresponding to $\sim$ 1 arcsecond resolution at the distance of the
observed galaxy NGC 3906. We varied the bin size in the analysis and found the effect to be minimal. 
The density map is then cut into logarithmically spaced intensity contours.  Each contour is fit with ellipses. 
The outer ellipse serves the stellar disk, while the innermost ellipse with eccentricity greater than 0.5 outlines the bar \citep{Yozin:2014wc}. Various methods have been used to identify the bar in our simulations; however, this approach produces the best fits to visual inspection, and allows for comparisons with previous analyses. 
The bar center has also been measured adopting the brightest pixel following the suggestion of \citet{vanderMarel:2014bi}, and we found that the bar center estimated in this way agrees with the values obtained with the ellipse method, with the uncertainties estimated to be within the numerical resolution. 

To find the disk and bar centers, we fit ellipses to the isodensity contours of our primary galaxy.
The disk center is probed by our outermost ellipse fit, which we set at a density of 0.5 \msun\ $pc^{-2}$.
The ellipse that fits this outer isodensity contour has a typical radius of 10 kpc. We tested the
  effect of measuring the offset from the dynamical center using a higher
  density cutoff of 10.7 \msun\ $pc^{-2}$. This density probes an inner
  disk with typical radii of 5 kpc. The inner disk shows similar qualitative
  results and a similar time evolution, but with reduced amplitude.

 The galaxy dynamical center is defined as the location where the galaxy
potential well is deepest. We measure it by taking the center of mass of the 100 particles with the most
negative potentials. We also measured the center of the dark matter alone by
repeating the same procedure, but using only dark matter particles. Throughout
the paper the bar and disk offsets refer to the displacement of that component with the galaxy dynamical center, but
the implications of using the halo center instead are minimal and will be
briefly discussed in \autoref{subsec:offsets}. The in-plane separation between the disk and bar centers is used as a measure of the bar-disk offset (what is traditionally called the ``offset bar''). 

An example of the photometric fitting approach applied to the stellar primary galaxy during the gravitational encounter with a companion is illustrated in \autoref{fig:combinedMaps} (top row), which shows the projected surface density and isodensity ellipses superimposed (orange lines).   
The inner ellipse (yellow line) identifies the stellar bar, and the outer ellipse (black line) fit the outer stellar disk. The panels are labeled by time, with the initial time of the simulation set to the time when the companion is first introduced and 
$t$ = 0 as the time when the perturber hits the disk of the primary galaxy.

\begin{figure*}[t] %  figure placement: here, top, bottom, or page
   \centering
   \includegraphics[width=7in]{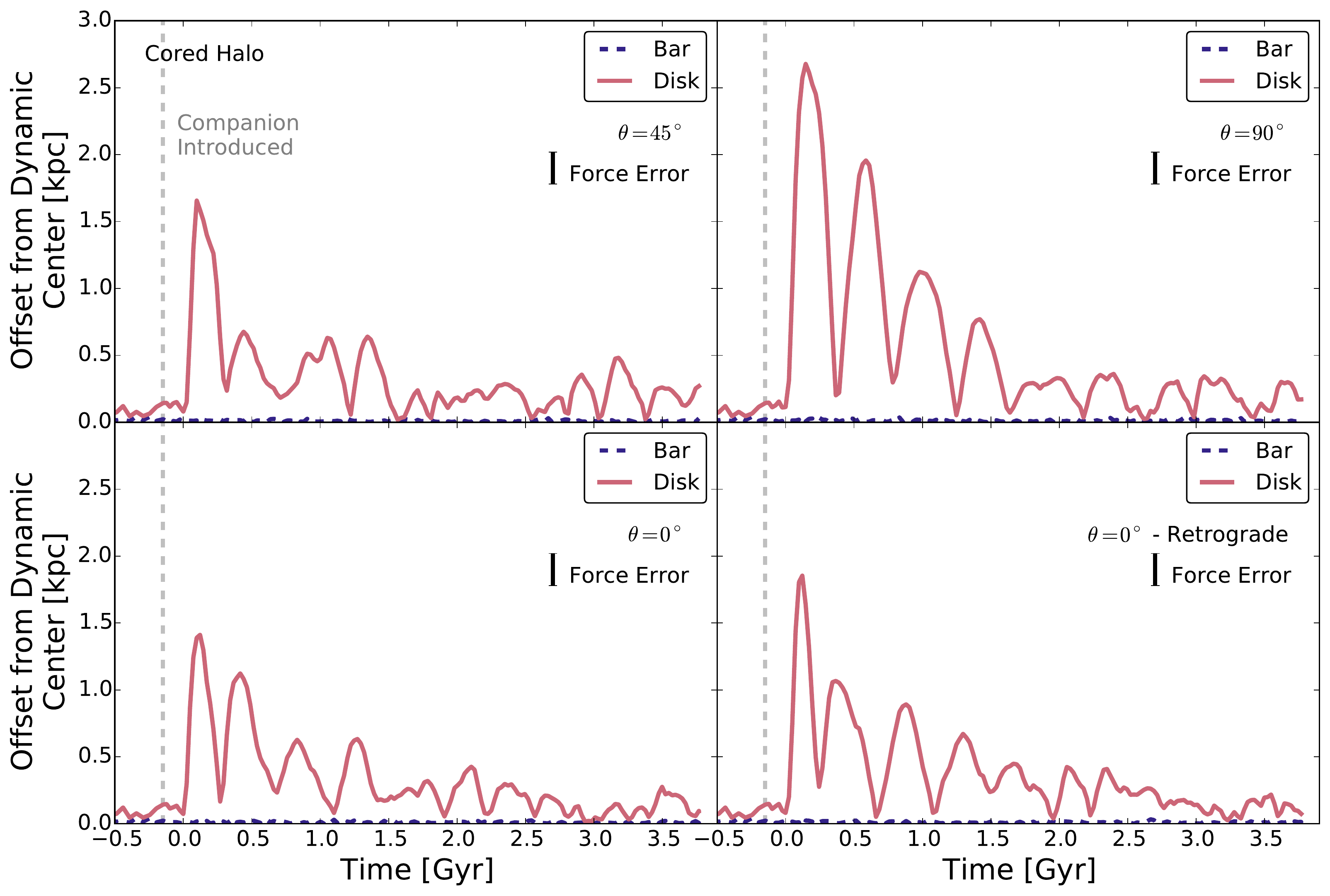} 
   \caption{Time evolution of offsets between the photometric center of the disk (solid
     red line) and bar (dotted blue line) and the dynamical center in the
     cored halo model for different orbital inclination angles. The size of the error bar is twice the typical force softening length $h_s$ =120 pc. The time of closest approach is set at 0 Gyr and we include 0.5 Gyr of evolution before the introduction of a companion. The gray vertical line indicates when the companion was introduced. \emph{Top left}: $\theta$=45$^{\circ}$. \emph{Top right}: $\theta$=90$^{\circ}$. \emph{Bottom left}: $\theta$=0$^{\circ}$. \emph{Bottom right}: $\theta$=0$^{\circ}$ (retrograde collision).}
   \label{fig:bar-disk}
\end{figure*}

\subsection{Quantifying the disk lopsidedness}
\label{subsec:fourier_methods}

In our simulations the asymmetries of the stellar disk and the lopsidedness are
determined by projecting  the  disk  face-on  and  measuring
its  stellar  surface  density, $\Sigma (r, \phi)$, with $(r, \phi)$  being  
polar  coordinates  in  the  disk  plane.    If  the  surface  brightness
distribution is invariant for a rotation of $2 \pi/m$ rad, so that $\Sigma
(r, \phi+2 \pi/m)=\Sigma(r, \phi)$, then the galaxy has
$m$-fold rotational symmetry and has $m$ arms and a bar for $m = 2$.
 
If the surface brightness is expressed as a Fourier series, then the disk
lopsidedness, the strength of the bar, and the amplitude of the spiral structure
can be measured from its  Fourier components, which are calculated as 

\begin{equation}
\begin{split}
\label{eq:fourier}
\Sigma_{mc}(r) = 2\langle \Sigma(r,\phi) \cos{m\phi} \rangle \\
\Sigma_{ms}(r) = 2\langle \Sigma(r,\phi) \sin{m\phi} \rangle \\
\textrm{for m = 1, 2, ..., }\infty \\
\Sigma_0 = 2\langle \Sigma(r,\phi) \rangle
\end{split}
\end{equation}
The amplitude of each Fourier component is calculated relative to $\Sigma_0$ and can take values between 0 and 1:
\begin{equation}
A_m = \frac{\sqrt{\Sigma_{mc}^2+\Sigma_{ms}^2}}{\Sigma_0}.
\end{equation}
The  face-on disk is divided into concentric annuli, which are further divided into azimuthal bins.  The mass
from  the  star  particles  is  then  assigned  to  these  bins,  making it
possible to compute the stellar surface density $\Sigma(r,\phi)$. Fourier components of the surface density are then calculated according to \autoref{eq:fourier}. 

Following \citet{Zaritsky:1997ev} and
\citet{Zaritsky:2013dj}, we then measure the average of the Fourier components' 
amplitudes $<A_m>$ (for $m$ = 1 and $m$ = 2) between 1.5 and 2.5 scale lengths (2.2 -
3.7 kpc for the disk scale length of 1.4 kpc). 
We repeat this measurement during the time of the interaction between the
primary galaxy and the companion to track the growth of asymmetries and the strength of the bar.

\subsection{Gas disk velocity fields}
\label{subsec:vel}

A weighted two-dimensional velocity map for the gas is inferred in our simulations   
projecting the galaxy as if it had been observed at a certain inclination and then taking the mass-weighted average of the radial
velocity components of the gas particles. An example of two-dimensional velocity maps is illustrated 
in the bottom row of \autoref{fig:combinedMaps}, where the galaxy is observed with an inclination of 45$^{\circ}$ 
starting before the companion is introduced and following several gigayears  of evolution.

%--------------------------
%RESULTS
%--------------------------
\section{Results}
\label{sec:results}

\subsection{Stellar disk dynamical response}
\label{subsec:offsets}

We investigate perturbations induced in the stellar disk of the
primary galaxy by a recent encounter with a companion exploring
four different orbital configurations: a coplanar (prograde and retrograde) 
an inclined orbit of 45$^{\circ}$ and 90$^{\circ}$, respectively.

\autoref{fig:combinedMaps} shows the live disk displayed
face-on during the 45$^{\circ}$ direct encounter with the companion. 
In this numerical experiment the disk galaxy is embedded in a
cored dark matter halo. There are observed asymmetries in the mass
distribution of the stellar disk noticeable in the outer parts, which appear 
shortly after the encounter. 

Next, we measured the dynamical center, the bar, and the disk photometric center.  
Concentric ellipses are superimposed to the disk
surface density as shown in \autoref{fig:combinedMaps}. 
The inner ellipse--marked in yellow--outlines the bar, and the
ellipse colored in black matches the outer disk density. 
The time of closest approach happens roughly 0.15 Gyr from 
the beginning of the simulation. The companion galaxy passes through the primary galaxy quickly ($\sim$0.1 Gyr within 20 kpc of the primary galaxy) and is within the virial radius of the primary galaxy for $<$0.5 Gyr.

\begin{figure}[htbp] %  figure placement: here, top, bottom, or page
   \centering
   \includegraphics[width=3.5in]{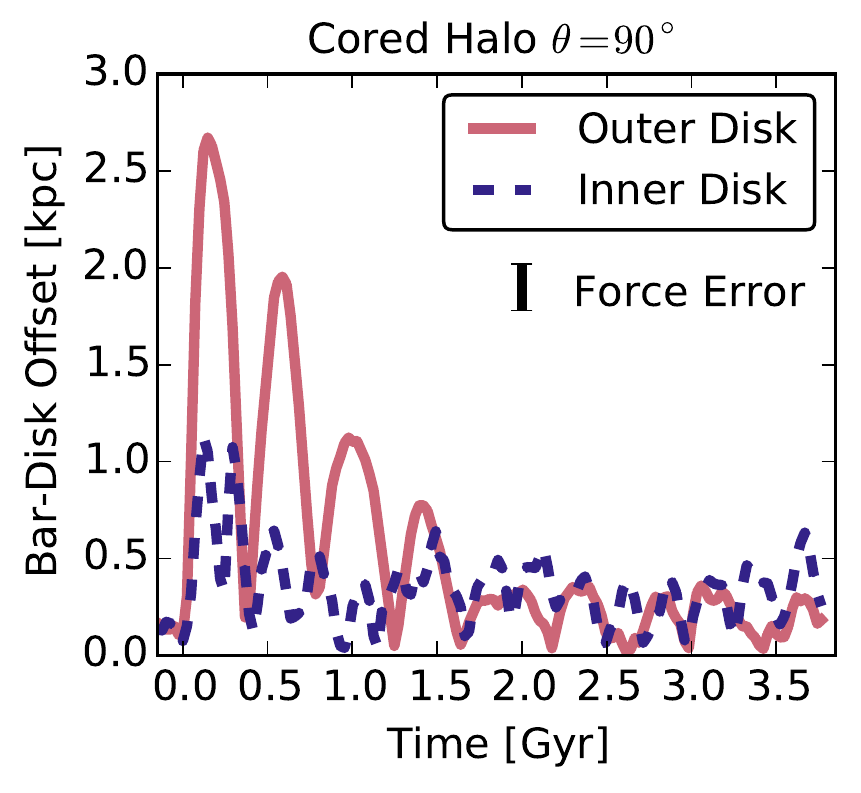} 
   \caption{Time evolution of the displacement between the bar and disk centers shown for the $\theta$=90$^{\circ}$ orbit in the cored halo. We show the effect of using a higher density cutoff which probes the inner disk, with a typical radius of 5kpc. The inner disk cutoff is 10.7 \msun\ $pc^{-2}$ while the outer disk cutoff is 0.5 \msun\ $pc^{-2}$. This example is the most extreme difference of any of our simulations, but still shows a strong offset between the measured disk and bar centers directly after impact.}   
   \label{fig:inner-outer-disk}
\end{figure}

We repeated the experiment with different sets of models varying the
orbital inclination angle. We measure the displacements between the dynamical center, the
stellar disk, and the bar. \autoref{fig:bar-disk} quantifies
the displacement of the bar and the stellar disk from the dynamical center
and displays it as a function of time for encounters run with
different orbital inclinations. We also ran our experiments on the isolated disks to measure the intrinsic scatter. In general, this scatter is quite low and well behaved.

After the encounter with the companion (labeled as $t$ = 0), the stellar disk
responds to the gravitational perturbation by becoming asymmetric in
its mass distribution (see the third panel in \autoref{fig:combinedMaps}).
We note first that the measured bar center
is always coincident with the dynamical center.  Hence, there is no
evidence of off-center bars for any orbital configuration. The stellar
disk, however, is measured to be 1.5-2.5 kpc shifted from the dynamic
center of each primary galaxy.  This readjustment of the disk occurs over a
period of approximately 2 Gyr, with small offsets of $\sim$0.5 kpc persisting for another 2 Gyr. 
During this time, the disk center will be displaced with respect to the bar center up to 2.5 kpc for encounters
with orbital inclination angles of 90$^{\circ}$ (the largest of the
observed shifts). 

The displacement of the primary galaxy's stellar disk is caused by the companion galaxy
passing though the disk. This fly-by creates strong asymmetries by scattering
stars to large radii on one side of the galaxy. This effect is more severe when
the impact occurs in the 90$^{\circ}$ orbit, producing the strongest disk
response and largest displacements from the dynamical center. In some cases, the asymmetric material appears as a single spiral arm.

\begin{figure*}[t] %  figure placement: here, top, bottom, or page
   \centering
   \includegraphics[width=7in]{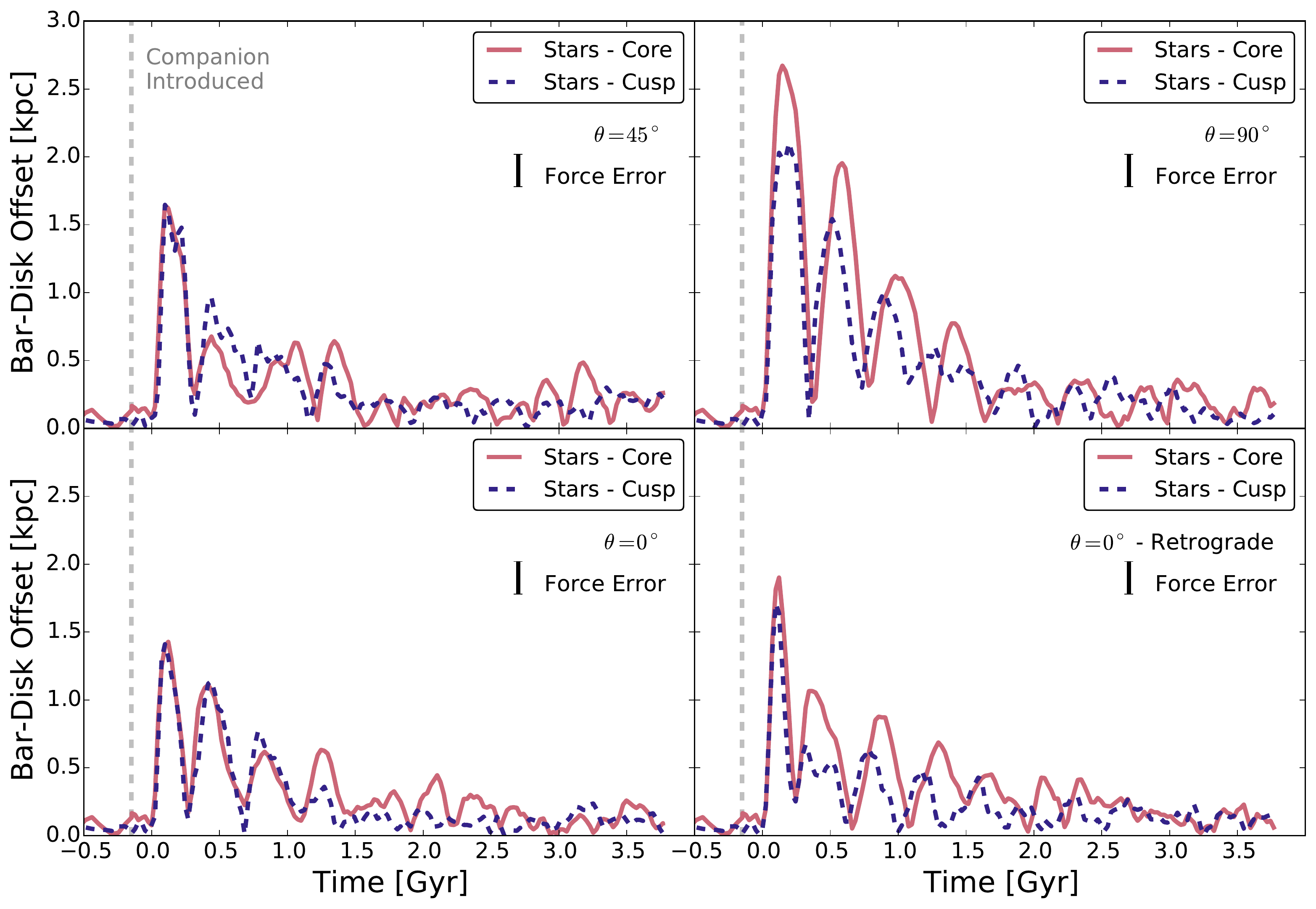} 
   \caption{Time evolution of the displacement between the bar and disk centers shown for the same
     four orbital orientations as in \autoref{fig:bar-disk}, and for both halo types. We compare models with a
     cored halo density profile (red lines) to models with a Hernquist inner-density profile for the dark
     halo (blue dashed-lines). The time of closest approach is set at 0 Gyr, and we include 0.5 Gyr of evolution before the introduction of a companion.}   
   \label{fig:core-cusp}
\end{figure*}

The disk distortions and asymmetries persist for
almost 2 Gyr, the time that it takes for the disk to be
recentered, and well after the perturber has passed. The response of the stellar disk appears to oscillate with time as it reduces amplitude, as also noted in \citet{Yozin:2014wc}. The primary offset is the most prominent. The subsequent offsets have amplitudes that decrease strongly with the time since the interaction and soon reach values close to our spatial resolution limits and do not appear to be correlated with disk rotation or other disk properties.

We also tested the offset using the center of the dark matter halo instead of the dynamical center. The dark matter halo center is aligned with the dynamical center before the interaction but becomes offset from both the dynamical center and the bar during the passage of the companion. After the interaction ends, the halo and dynamical centers coincide again and align with the bar.

During testing of our orbital configurations, we made runs with various impact parameters. 
We took care to hold the impact parameter constant across all simulations with different configurations. 
Decreasing the impact parameter increased the observed bar-disk offset and measured asymmetries.  

As a final test of these results, we experimented with fitting an ellipse to a higher-density region of the disk. This ellipse probes the inner disk of the primary galaxy (defined by a density cutoff of 10.7 \msun\ $pc^{-2}$), with a typical radius of 5kpc (vs. our outer disk cutoff of 0.5 \msun\ $pc^{-2}$, which has a typical radius of 10 kpc). As an example in \autoref{fig:inner-outer-disk}, we show only the most different result between the inner and outer disk which occurs for the $\theta$=90$^{\circ}$ orbit in the cored halo. Although the qualitative results are the same, with an offset directly after the encounter followed by a gradual damping, the amplitude of the effect is strongly reduced when using the center of the inner disk. 

\subsection{Dependence on the shape of halo density profile}
\label{subsubsec:corecusp}

There is some evidence that dwarf disk galaxies do show slowly rising
rotation curves that would be consistent with having underlying
halos with shallow inner-density profiles (\citeauthor{2001ApJ...552L..23D}, \citeyear{2001ApJ...552L..23D}; \citeauthor{Oh:2011jd}, \citeyear{Oh:2011jd}; \citeauthor{2002sgdh.conf..186M}, \citeyear{2002sgdh.conf..186M}; but see also \citeauthor{Oman:2015gb}, \citeyear{Oman:2015gb}). Indeed, if Magellanic disk galaxies are embedded in halos with less concentrated mass distributions in the inner
regions, then we may expect an enhancement in the amplitude of the disk
oscillations around the dynamical center. If off-centered bars are a real phenomenon, then it is plausible to posit that
these oscillations would cause the bar to remain more offset, and for longer, than in a galaxy with a cosmological dark matter halo. To test whether this was driving the offsets seen in the previous section, 
we also set up models for the Magellanic disk galaxy with a cusped Hernquist dark matter density profile in the inner parts.

\autoref{fig:core-cusp} illustrates the outcome. Different panels display
the time evolution of the disk and bar offset for encounters (the
in-plane separation between the measured bar and disk centers) with
different orbital configurations. The largest
differences are seen in the 90$^{\circ}$ orbit, where the maximum (average)  
displacement is 0.6 (2.7) kpc for the cored profile, compared to 0.5 (2.1) kpc for the cusped profile.
The separation measured for the co-planar retrograde 0$^{\circ}$ case consists
in the maximum offset of 0.4 (1.9) kpc for the cored profile, as compared to
the displacement of 0.3 (1.7) kpc measured for the cusped profile.

The 45$^{\circ}$ inclined orbit shows a slightly stronger offset for the
model with the cusped halo, but this difference and the difference seen 
in the co-planar prograde ($\theta$=0$^{\circ}$) orbit are both close to the spatial resolution of our models.
These results support the conclusion that the bar-disk offsets
are driven  by asymmetries in the outer part of the disk.

\begin{figure*}[htbp] %  figure placement: here, top, bottom, or page
   \centering
   \includegraphics[width=7in]{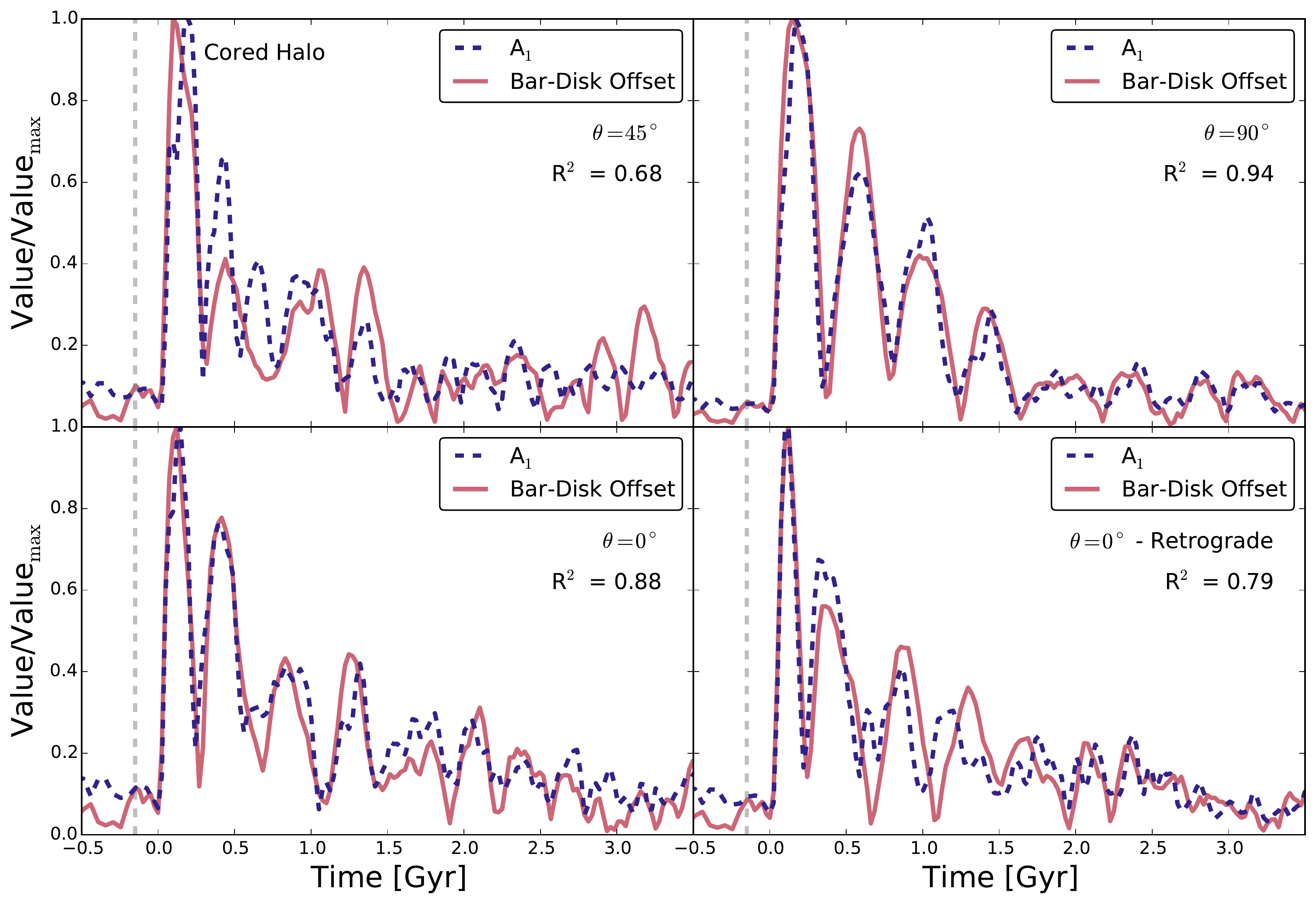} 
   \caption{Time evolution of the asymmetry of the stellar disk in the cored
     halo compared with the bar-disk offset for the same
     four orbital orientations as in \autoref{fig:bar-disk}. The two quantities have been normalized to their maximum value over the full time range. We show the R$^2$ value in each panel, quantifying the degree to which the two time series coincide.}
   \label{fig:fourier}
\end{figure*}

\subsection{Stellar disk lopsidedness}
\label{sec:fourier_results}

If offsets are the results of an asymmetric disk in the outer parts, then we expect this
to appear  in the Fourier components analysis.  To test this the Fourier components 
of the stellar disk of the primary galaxy have been measured as a function of time, during
the interaction with the companion and long after the encounter ended. Our galaxies begin with strong $A_2$ amplitudes indicating symmetric stellar features, and low $A_1$ amplitudes. 

The outcome is displayed in \autoref{fig:fourier} for the primary galaxy with a cored halo.
The results for the Hernquist halo are qualitatively similar. The four panels show the $A_1$ and bar-disk offset normalized to their maximum value in the four orbital configurations.
 
We find that $A_1$ becomes pronounced during the peaks of the observed bar-disk offset. These peaks line up remarkably well for $\sim$2 Gyr after the encounter, at which point the offsets subside. We can quantify the degree to which the two time series coincide by measuring the R$^2$ coefficient:
\begin{equation}
R^2 = 1 - \frac{\sum_{t=0}^{4} (o(t) - \overline{o(t)})^2}{\sum_{t=0}^{4}(o(t) - f(t))^2},
\end{equation}
where, in this case, $o(t)$ is the measured bar-disk offset at time $t$ normalized to its maximum value and $f(t)$ is the $A_1$ fourier amplitude at time $t$ normalized to its maximum value. An R$^2$ value of 1 would indicate that the value of the $A_1$ mode can explain 100\% of the bar-disk offsets, while an R$^2$ value of 0 would indicate that the the $A_1$ mode cannot explain any of the bar-disk offsets. It is important to note, however, that this value merely describes the similarity of the two datasets and cannot prescribe any physical causation one way or the other. The lowest R$^2$ coefficient is found for the $\theta = 45^{\circ}$ configuration with R$^2$ = 0.68. The highest value is found for the $\theta = 90^{\circ}$ with R$^2$ = 0.94. 

In every orbital configuration modeled except when $\theta = 0^{\circ}$ in a retrograde orbit, $A_1$  
becomes the dominant mode directly after encounter and remains elevated from its initial amplitude for the entire 4 Gyr. 
For the $\theta = 0^{\circ}$ retrograde orbit $A_1$ is dominant for a short
period of time, and settles within 3 Gyr. Using the criteria of \citet{Bournaud:2005ey} and \citet{Mapelli:2008fx} (A$_1 \ge 0.05$), the galaxies showing evidence for bar-disk offsets all appear asymmetric during these periods, a result also found in \citet{Athanassoula:1997tx} and \citet{Berentzen:2003dw}. The average bar amplitude in our cored halo shows a modest decrease after the interaction.

Although qualitatively similar (both show higher $A_2$ amplitudes before the interaction with the companion 
and strong $A_1$ amplitudes during the tidal interaction), 
the results for the Hernquist halo have a number of notable differences. The Hernquist Fourier component is 
A$_2< 0.2$ before interaction, even though the galaxy appears strongly barred according to our visual inspection. 
During the tidal interaction, both $A_1$ and $A_2$ dominate over the higher Fourier components. 

Because of this, we can conclude that the shift of the disk center away from the dynamical center is due to the extended tidal material created after the interaction with the smaller companion. The method of measuring disk centers using fits to isodensity contours appears to be sensitive to larger asymmetric features on the edge of the disk. This conclusion is further supported by the fact that the typical radius measured by the outer isodensity contours increases during the encounter with the companion.

%--------------------------
%DISCUSSION
%--------------------------

\subsection{Gas disk dynamical response}
\label{sec:discussion}

\begin{figure}[htbp] %  figure placement: here, top, bottom, or page
   \centering
   \includegraphics[width=3.5in]{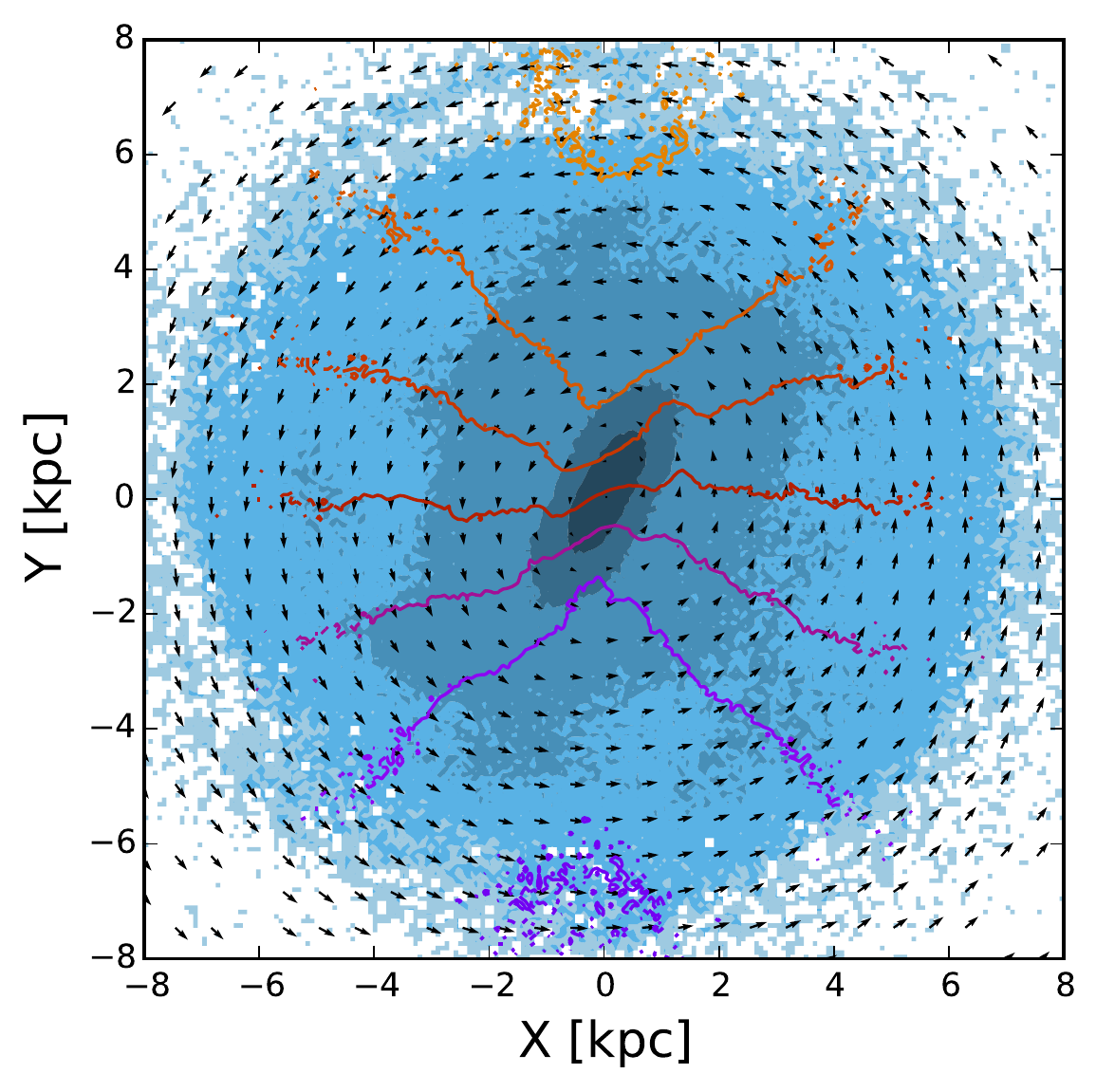} 
   \caption{Velocity fields of the stellar and gas components overlaid on the
     stellar density distribution from the primary galaxy just before the
     encounter with the companion galaxy occurs. This model assumes a cored
     halo profile for the primary galaxy. The blue filled contours in the
     background display the stellar density distribution as in the first row
     of \autoref{fig:combinedMaps}. 
    The colored contours show the gaseous velocity fields as on the bottom row
    of \autoref{fig:combinedMaps}. 
    The arrows show the average stellar motion in the x-y plane in each
    pixel. Note that only 1/8 of the arrows are plotted here to improve clarity.}
   \label{fig:velfield}
\end{figure}

In the simulations described in the previous sections, we showed that
 the bar accurately reflects the dynamical center, while the stellar disk exhibits a prominent offset. 
\citet{Wilcots:2004gv}, however, find that the majority of the observed Magellanic galaxies have \HI\ profiles no more or less asymmetric than other late type spirals found in \citet{Matthews:2002gj}.
To test the effects of the gravitational encounter with a companion on the gas distribution
of the primary galaxy, we measured the centers of the simulated gas velocity fields
 adopting a 2D nonaxisymmetric disk model as implemented  
the code \textsc{DiskFit} \citep{Spekkens:2007bf, Sellwood:2010fu}. This method uses a 
nonaxisymmetric model to fit the kinematic data of the gas particles within a radius of 14 kpc
and compares with previous methods.

To further test how the asymmetries appear in the velocity distributions, we measure the average in-plane 
velocity of the disk stars in the each pixel of our surface density map. 
We demonstrate this method by superposing the velocity vectors on the surface density in \autoref{fig:velfield}. 
This method is similar to the one applied to the LMC stellar disk using proper-motion rotation fields by \citet{vanderMarel:2014bi}. 

\begin{figure*}[t] %  figure placement: here, top, bottom, or page
   \centering
   \includegraphics[width=7in]{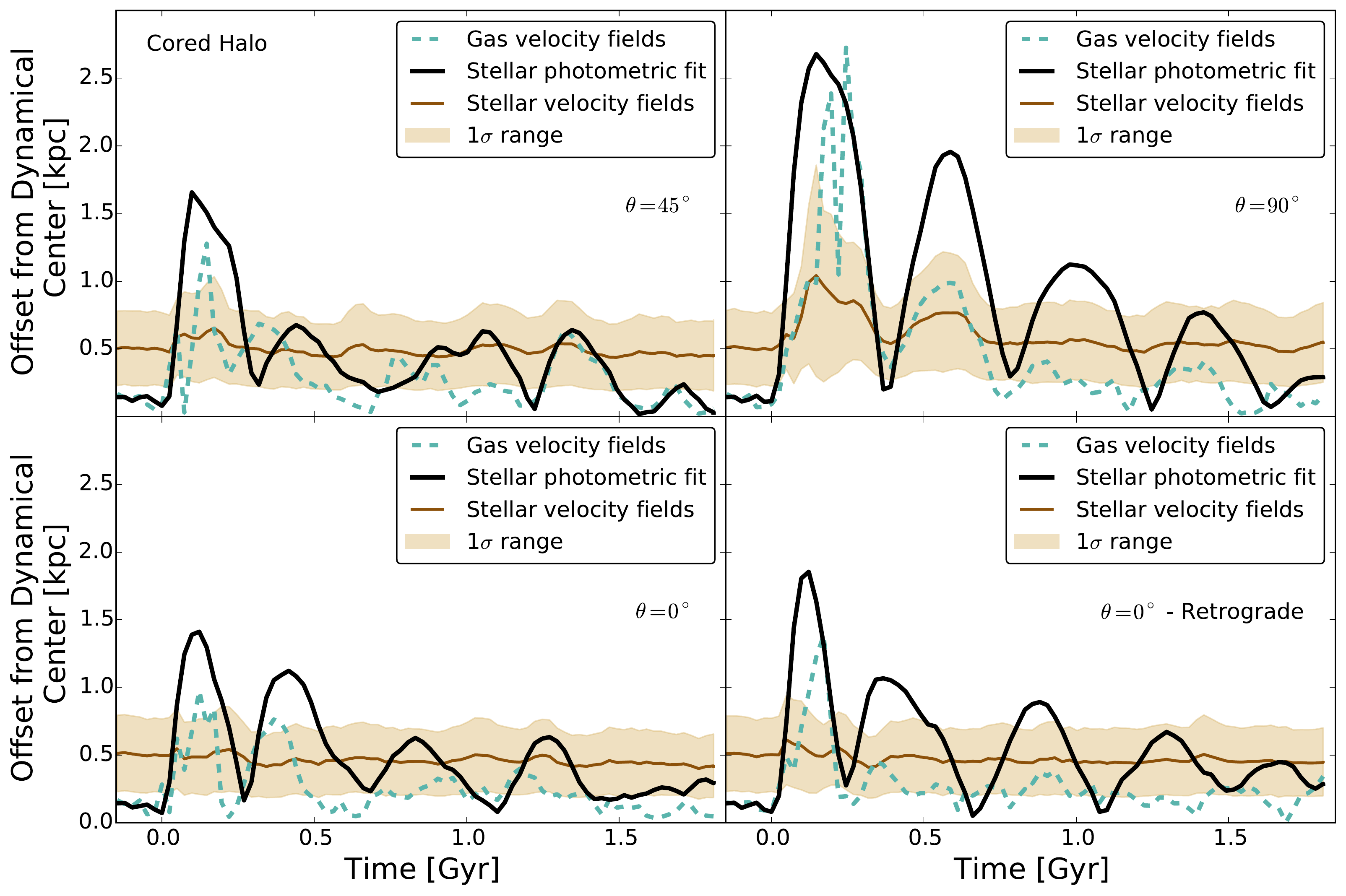} 
   \caption{Time sequence of offset between gas and stellar disk and the dynamical center as measured 
by using different methods and for encounters with different orbital configurations. The primary galaxy
has a cored halo profile. 
Centers of the disk measured from the photometric ellipse fitting are shown 
(thick black line). Centers of the gaseous disk measured using 2D tilted ring fitting of the gaseous velocity field by DiskFit are also displayed (dotted blue line) and compared to 
centers of the stellar disk as measured using 
the average in-plane velocity of the stars (thin brown line).  
The solid line shows the average of all 10,000 trials (see discussion in \autoref{subsec:lmc}), 
and the filled region shows the standard deviation of all the trials.}
   \label{fig:velcenters}
\end{figure*}

The centers obtained with this method at each time 
are compared to the dynamical center as previously measured 
and displayed in \autoref{fig:velcenters}. The stellar photometric center obtained by fitting ellipses 
to the stellar disk and bar is shown (solid black line), as compared to the
the gas velocity field center measured by DiskFit (dotted blue line),
 and with the stellar velocity field center 
described above (solid brown line).

First, we notice that the velocity field center of the stellar disk is never coincident with the dynamical 
center, defined as the location of the deepest potential well, but it is always displaced on average 0.5 kpc, and with a maximum displacement of $\sim$1 kpc when $\theta = 90^{\circ}$.

Our findings are in agreement with Magellanic spiral galaxies observed in the field.  
In these systems the dynamical center, usually assumed to be traced  
from \HI\ rotation curves, is coincident with the center of the bar and offset with respect 
to the photometric center of the stellar disk,  as shown in the Magellanic galaxy NGC 4027 \citep{Phookun:1992dc} and in NGC 3906 \citep{2011ApJS..194...36W, deSwardt:2015fk}. Similar behavior was found by \citet{Matthews:2002gj} 
studying  a sample of typical late-type spiral galaxies. In particular, the latter work showed 
that the global \HI\ profiles were generally symmetric and traced the center of the galaxy's potential 
even when the optical image or kinematic data showed asymmetry.

We also observe that the center location found for the gas velocity field is different than the center location of the stellar disk as measured by the the outer ellipse. Minor offsets between the gas and stars, like the one we find here, have been seen before in spiral arms created during encounters with lower-mass companions \citep{Pettitt:2016iv} and along the leading edges of strong central bars \citep{1983IAUS..100..215P, 1992MNRAS.259..345A, Sheth:2002gg}.

\subsection{Determination of the LMC dynamical center}
\label{subsec:lmc}

There have been some questions raised about whether the bar in the LMC is a real dynamical feature. 
\citet{Zaritsky:2004iw} argues that it might be a triaxial bulge viewed edge-on, but \citet{Subramaniam:2009vg} show that the bar is clearly part of the disk and may be influencing the gas \citep{Indu:2014vh}.

The bar is thought  to be offset from the disk center \citep{vanderMarel:2001cr} and from the 
dynamical center as measured from \HI\ kinematic tracers \citep{Luks:1992wk, Kim:1998eh, vanderMarel:2001cr}. Interestingly, the bar and outer disk are both consistent, within the estimated uncertainties, 
with the dynamical center as traced by the line-of-sight velocities of carbon stars 
\citep{2002AJ....124.2639V}. 

Using stellar proper-motion fields, \citet{vanderMarel:2014bi} determined a more accurate dynamic center that is coincident with the \HI\ center. 
According to our results such an offset can occur very soon after the encounter with a companion, 
as was likely the case with the LMC $\sim$200 Myr ago \citep{Besla:2012jc} when it collided with the SMC.

Following \citet{vanderMarel:2014bi} (their Fig. 2),  we mapped the center of our simulated 
two-dimensional stellar velocity map of the LMC analog primary galaxy after the collision with the SMC analog.
In order to mock the observed stellar velocity map, we chose a sample of 22 fields 
(the number of fields used in the analysis of \citet{vanderMarel:2014bi}) 
within 5 kpc of the disk center of mass and fit the rotation curve:
 \begin{equation}
 V(r) = V_0 \times \rm{min}\left[r/R_0, 1\right],
 \end{equation}
 with $r = \sqrt{(x-x_0)^2+(y-y_0)^2}$.
Here ($x_0$, $y_0$) are center coordinates, $V_0$ is the maximum velocity, and $R_0$ is the turnover radius. We repeat this sampling process 10,000 times at each time step, mocking the process of observing a random 
population of stars of the galaxy as in the observations. We measure in this way at each time a systematic  
offset from the true dynamical center, traced in our simulations by the location where the potential is 
deepest. The displacement is amplified during the collision as shown in \autoref{fig:velcenters}
(solid brown line).  

Our results predict that the halo center of the LMC is coincident with the bar center and that the \HI\ centers will realign with this center over the next few hundred million years. 
If confirmed, this finding suggests that the bar center should be assumed as dynamical center instead
of the \HI\ center that has been previously used \citep{Kallivayalil:2013vq}. 
This choice would increase the north component direction of the proper-motion
measurements, $\mu_N$, from the current value $\mu_N=0.229 \pm 0.047$ to values close to
the previous estimate of $\mu_N=0.34$ \citep{2002AJ....124.2639V}. 
The north component of the proper-motion of the LMC controls the location
of the orbit when projected on the plane of the sky. This correction will increase the offset between the LMC's orbit and the position of the Magellanic Stream, an issue still unsolved in the studies on the origin of the Stream.

%--------------------------
%CONCLUSIONS
%--------------------------
\section{Discussion and Conclusions}
\label{sec:conclusions}

We have investigated the perturbations induced in the stellar and gas disc
and the bar of a Magellanic-type spiral galaxy by a fly-by with a companion.
A set of numerical experiments have been studied varying the angle between the orbital plane and the equatorial plane of   the primary galaxy. The dynamical response of the disk has also been analyzed for a cored or Hernquist dark halo profile for the Magellanic-type galaxy. The results can be summarized as follows:\\

\begin{enumerate}
\item The bar center is always coincident with the dynamical center, suggesting that, 
contrary to common belief, the bar is never displaced in Magellanic spiral galaxies. Instead, the stellar disk 
is measured to be shifted from the dynamic center and the bar of the 
primary galaxy by, at most, 1.5-2.5 kpc, depending on the details of the 
encounter with the companion.
Thus, the observed displacements should, on average, be well below this value, in good agreement with observations \citep[e.g.][]{1980SSRv...27...35F}. The disk distortions and asymmetries persist for almost
2 Gyr, the time that it takes for the disk to be recentered,
and well after the interaction with the companion ended. 
The disk sloshing around the dynamical center reduces in amplitude with time. 

\item Disk asymmetries are slightly more pronounced in Magellanic galaxies
with a cored halo profile as compared to models with a cuspy halo profile
for the dark matter, but the differences are modest. The largest response of the disk
of the primary galaxy tidally induced by the passage of the companion occurs
for smaller impact parameters and depends on the orientation of the orbital plane with respect to the primary galaxy's equatorial plane.

\item The  gas disk also sloshes similarly to the stellar disk during the gravitational 
interaction with the companion, but with a modest
amplitude, and it tends to recenter after a short time due to the dissipative nature of the gas.

\item These results, when applied to the LMC -- the prototype of the Magellanic spiral galaxies --
suggest that the dynamical center should reside in the bar center instead of the \HI\ center
usually assumed in previous works. This choice would imply a change in the north component 
of the LMC proper-motion estimates, perhaps pointing to interesting implications for the offset
between the LMC's orbit and the position of the Magellanic Stream. However, the little observational
evidence of a bar in the \HI\ gas distribution \citep{StaveleySmith:2003bl} remains a challenge 
for theoretical models.

\end{enumerate}
Our results might be of interest for interpreting the recent discoveries of bulgeless 
galaxies with off-center `nuclear' clusters (Gallagher, J.S. 2016 private communication).  
This is a larger class of objects with stellar disks showing mismatched photometric centers. 
According to our results, their dynamical centers are likely to reside in the nuclear clusters,
while the irregularities and asymmetries of the stellar disks should be interpreted as the outcome of tidally-induced distortions.

\acknowledgments

Support for this research was provided by the University of Wisconsin - Madison 
Office of the Vice Chancellor for Research and
Graduate Education with funding from the Wisconsin
Alumni Research Foundation. This research is also funded by NSF Grant No AST-1211258 and
ATP NASA Grant No NNX144AP53G. ED gratefully
acknowledges the support of the Alfred P. Sloan Foundation. 
EA acknowledges financial support from the EU Programme FP7/2007-2013/, 
under REA grant PITN-GA-2011-289313 and  from CNES (Centre National d'Etudes Spatiales, France).
We express our appreciation to the Aspen Center for Physics for their hospitality, funded by
the NSF under Grant No. PHYS-1066293. Simulations have been run on the High Performance Computing
cluster provided by the Advanced Computing Infrastructure (ACI) and Center for High Throughput Computing
(CHTC) at the University of Wisconsin.
This research made use of Astropy\footnote{http://www.astropy.org/}, a community-developed core Python
package for Astronomy (Astropy Collaboration, 2013). This research made use of the
Pathos\footnote{http://trac.mystic.cacr.caltech.edu/project/pathos}
multiprocessing library \citep{McKerns:2012wd}. This research has made
use of the NASA/IPAC Extragalactic Database (NED), which is operated
by the Jet Propulsion Laboratory, California Institute of Technology,
under contract with the National Aeronautics and Space Administration,
and NASA's Astrophysics Data System. 

\bibliography{magellanic_offsets}
\end{document}